\documentclass[10pt,journal,compsoc]{IEEEtran}

\ifCLASSOPTIONcompsoc
  \usepackage[nocompress]{cite}
\else
  \usepackage{cite}
\fi

\ifCLASSINFOpdf
  \usepackage[pdftex]{graphicx}
\else
  \usepackage[dvips]{graphicx}
\fi

\usepackage{amsthm}
\usepackage{amsmath}
\usepackage{bm}
\usepackage{multirow}
\usepackage{enumitem}
\usepackage{amssymb}

\usepackage{url}

\begin{document}

\title{Enhancing Factorization Machines with Generalized Metric Learning}


\author{Yangyang Guo,
        Zhiyong Cheng,
        Jiazheng Jing,
        Yanpeng Lin,
        Liqiang Nie~\IEEEmembership{Senior Member,~IEEE,} \\
        Meng Wang~\IEEEmembership{Senior Member,~IEEE}

\IEEEcompsocitemizethanks{
\IEEEcompsocthanksitem Yangyang Guo is with Shandong Artificial Intelligence Institute, Qilu University of Technology (Shandong Academy of Sciences), China, and also with Shandong University, China. Email: guoyang.eric@gmail.com
\IEEEcompsocthanksitem Zhiyong Cheng is with Shandong Artificial Intelligence Institute, Qilu University of Technology (Shandong Academy of Sciences), China.
E-mail: jason.zy.cheng@gmail.com.
\IEEEcompsocthanksitem Jiazheng Jing and Liqiang Nie are with Shandong University, China. E-mail: \{jingjiazheng, nieliqiang\}@gmail.com.
\IEEEcompsocthanksitem Yanpeng Lin is with Mercari, Inc., Japan. E-mail: ypenglyn@mercari.com.
\IEEEcompsocthanksitem Meng Wang is with Hefei University of Technology, China. E-mail: eric.mengwang@gmail.com.
}
\thanks{The work was done when Yangyang Guo was an intern at Shandong Artificial Intelligence Institute, Qilu University of Technology (Shandong Academy of Sciences).}
\thanks{Zhiyong Cheng and Liqiang Nie are the corresponding authors.}
}

\IEEEtitleabstractindextext{%
\begin{abstract}
Factorization Machines (FMs) are effective in incorporating side information to overcome the cold-start and data sparsity problems in recommender systems. Traditional FMs adopt the inner product to model the second-order interactions between different attributes, which are represented via feature vectors. The problem is that the inner product violates the triangle inequality property of feature vectors. As a result, it cannot well capture fine-grained attribute interactions, resulting in sub-optimal performance. Recently, the Euclidean distance is exploited in FMs to replace the inner product and has delivered better performance. However, previous FM methods including the ones equipped with the Euclidean distance all focus on the attribute-level interaction modeling, ignoring the critical intrinsic feature correlations inside attributes. Thereby, they fail to model the complex and rich interactions exhibited in the real-world data. To tackle this problem, in this paper, we propose a FM framework equipped with generalized metric learning techniques to better capture these feature correlations. In particular, based on this framework, we present a Mahalanobis distance and a deep neural network (DNN) methods, which can effectively model the linear and non-linear correlations between features, respectively. Besides, we design an efficient approach for simplifying the model functions. Experiments on several benchmark datasets demonstrate that our proposed framework outperforms several state-of-the-art baselines by a large margin. Moreover, we collect a new large-scale dataset on second-hand trading to justify the effectiveness of our method over cold-start and data sparsity problems in recommender systems. \end{abstract}

}

\maketitle

%
\IEEEpeerreviewmaketitle

\IEEEraisesectionheading{\section{Introduction}\label{introduction}}
\IEEEPARstart{I}{t} is well-known that the performance of recommender systems is unstable under the cold-start and data sparsity settings~\cite{cold,cold_tkde}. The exploitation of side information provides an effective way to tackle these problems and many approaches have been developed over the years~\cite{amazon2}. Among these approaches, Factorization Machines (FMs)~\cite{fm,libfm} have gained more and more attention recently~\cite{nfm,afm,transfm}, which can easily leverage any side information (including user- and item-related) for recommendation.

FMs first map each attribute into a latent space and then concatenate the embedded vectors of attributes to a high-dimensional sparse vector\footnote{An Attribute can be a user ID, item ID, or other contextual attributes, e.g., user age. In this paper, an attribute is represented by a $k$-dimensional feature vector, indicating that $k$ features are used to describe an attribute in the embedded space.}. Particularly, FMs predict the target mainly by modeling all the second-order interactions among attribute using a factorized parametrization. Despite their promising performance, traditional FM methods suffer from two limitations: 1) the attribute interactions are modeled in a linear way (i.e., the predicted target is linear \textit{w.r.t.} each model parameter), which is insufficient for capturing the non-linear and complex inherent structure of real-world data~\cite{nfm}. And 2) FMs model the second-order interactions between attributes via the inner product of their factorized vectors. The problem is that the inner product does not satisfy the triangle inequality\footnote{It is defined as: ``The distance between two points cannot be larger than the sum of their distances from a third point.''~\cite{triangle}. Specifically, for real valued vectors $\bm{x}$, $\bm{y}$ and $\bm{z}$, triangle inequality requires meeting the condition that $d(\bm{y}, \bm{z}) \leq d(\bm{x}, \bm{y}) + d(\bm{x}, \bm{z})$.} in the vector space, which is crucial to model fine-grained relationships between attributes, resulting in sub-optimal performance~\cite{metriclearning,cml}.

\begin{figure}
  \centering
  \includegraphics[width=0.9\linewidth]{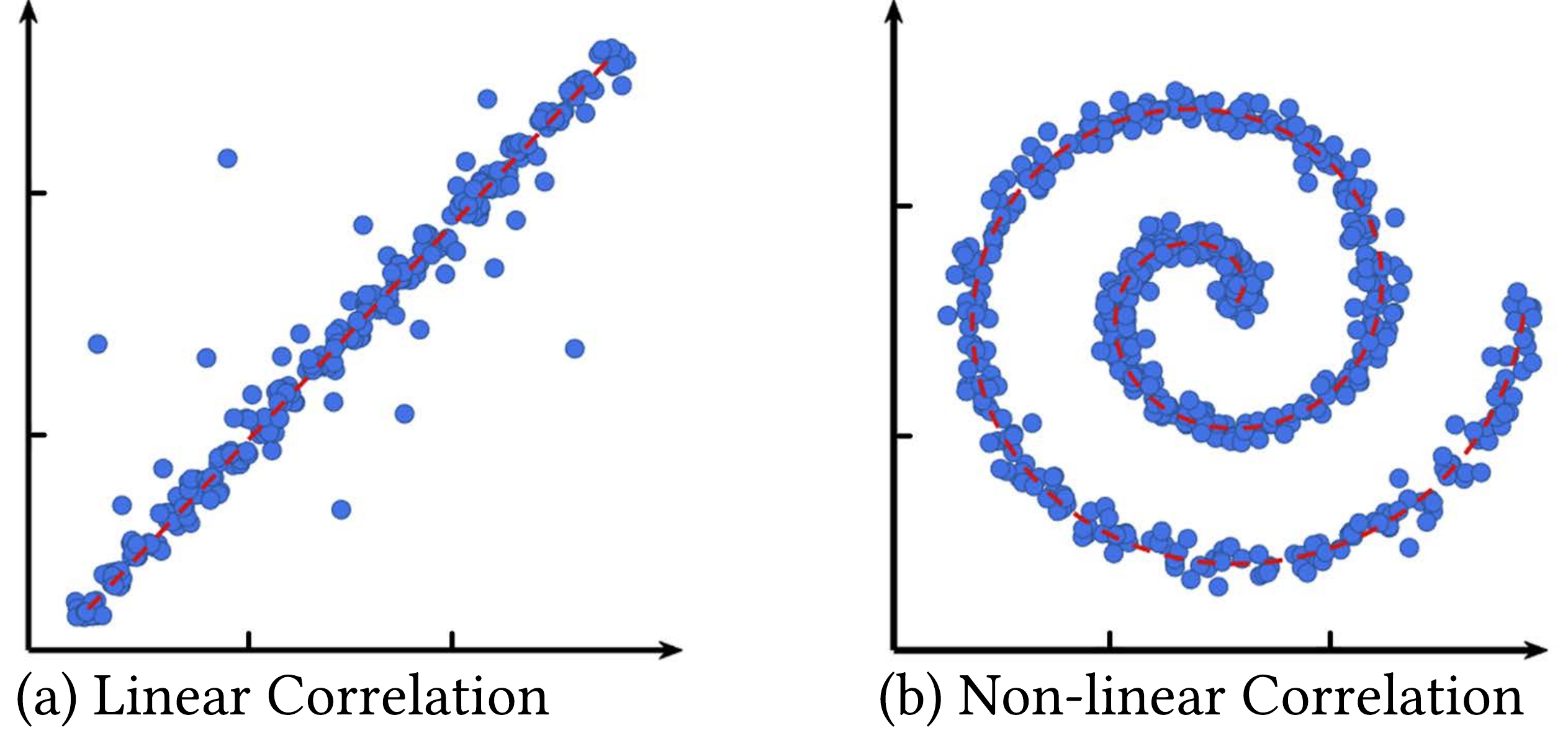}
  \vspace{-1em}
  \caption{An exemplar illustration of two kinds of feature correlations.}\label{fig:relation}
\end{figure}

To overcome the first problem, researchers have explored deep neural networks for non-linear transformations. For example, He et al.~\cite{nfm} developed a Bi-Interaction operation upon the factorized features, followed by several multi-layer perceptrons (MLPs) to capture the non-linear attribute interactions. However, this method is still unable to address the second problem, leading to inferior performance as observed in our experiments (see Section~\ref{res:all}). To tackle the second problem, Pasricha et al.~\cite{transfm} developed a new FM method, which replaces the inner product with a distance function (i.e., squared Euclidean distance) in FMs for recommendation. Specifically, they employed the squared Euclidean distance to estimate the similarity between each pair of feature vectors. It is worth noting that the Euclidean distance function computes the distance between each pair of the factorized feature vectors independently and then sums them up. It thus ignores the possible inherent correlations between different features of an attribute. Taking two typical examples as shown in Figure~\ref{fig:relation}, we map the attributes into a 2-D feature space. There are 1) a linear correlation in Figure~\ref{fig:relation}(a), indicating a positive correlation between two features, such as the \emph{acting performance} and \emph{actor popularity} for the attribute of a movie's \emph{leading actor}, and 2) a non-linear correlation in Figure~\ref{fig:relation}(b), e.g., the complex correlation between \emph{music rhythm} and \emph{music melody} for the attribute \emph{music elements}. When using the Euclidean distance function to compute the similarity between two features with a certain correlation, as the ones shown in Figure~\ref{fig:relation}, it often fails to capture such relationships between features. As a result, it is incapable of modeling the fine-grained feature interactions of attributes.

Motivated by the above observations, in this paper, we devise a novel FM framework equipped with generalized metric learning techniques (dubbed as GML-FM). Based on this framework, we study two different distance methods: the Mahalanobis distance and DNN-based distance methods. The Mahalanobis distance based method adopts a positive semi-definite matrix to project the features into a new space such that the features obey certain linear constraints. In this way, the linear correlations between features such as the ones in Figure~\ref{fig:relation}(a) can be captured by this matrix. To model the more complex correlations such as the one shown in Figure~\ref{fig:relation} (b), the DNN-based distance function is designed to capture the non-linear feature correlations, which can benefit from both the metric learning approaches and the strong representation capability of neural networks. Notice that the values of the traditional inner product can cover the whole real number space, while the values from distance functions are all non-negative, which will limit the representation capability of FMs. To tackle this problem, we introduce a learnable weight to the interactions of each attribute pair. This strategy can greatly enhance the performance of distance functions (see Section~\ref{res:ablation}).

Extensive experiments have been conducted on four public benchmark datasets, including three widely-used Amazon datasets~\cite{amazon2} and the MovieLens one~\cite{movielens}. Comparisons with several state-of-the-art methods demonstrate the effectiveness of our method. To further explore the superiority of our method over a variety of baselines under sparse settings, we collected a new large-scale dataset on second-hand trading from Mercari\footnote{https://mercari.com/.}, which is of high sparsity (i.e., most items are only purchased once), and contains rich side information (e.g., item condition, shipping duration). Experiments on this dataset also validate the effectiveness of our method.

In summary, our main contributions are four-fold:
\begin{itemize}
\item We propose a novel FM framework equipped with generalized metric learning techniques to effectively model the fine-grained feature interactions inside attributes. This framework can generalize the traditional inner product based and the recently proposed FMs with the Euclidean distance.
\item Based on the proposed method, we further design an effective solution to simplify the model equations and verify that our proposed method can be implemented in an efficient way.
\item We collect a new large-scale second-hand trading dataset to facilitate the study of the cold-start and sparsity problems in recommendation. To the best of our knowledge, this Mercari dataset is the largest second-hand trading dataset for recommendation in literature.
\item We conduct extensive experiments on three Amazon datasets, MovieLens dataset and two Mercari datasets to validate the effectiveness of the proposed method. Moreover, we have released the code to facilitate future research in this direction\footnote{\url{https://github.com/guoyang9/GML-FM}.}.
\end{itemize}

The rest of the paper is structured as follows. We define some preliminaries in Section~\ref{preliminary}. We then detail our framework and its simplification form in Section~\ref{model}. Experimental setup and result analysis are presented in Section~\ref{setup} and ~\ref{results}, respectively. In Section~\ref{related_work}, we briefly review the related work. We finally conclude our work and discuss the future directions in Section~\ref{conclusion}. 
\section{Preliminaries}\label{preliminary}
We first shortly recapitulate the key definition in literature, and then introduce the mainstream involvement of Factorization Machines. Both of these two are the building blocks for our proposed method.
\subsection{Metric Learning}
Given a data collection $\mathcal{V} = \{\bm{v}_1, \bm{v}_2, ..., \bm{v}_n\}$, where each data sample is over the input space $\mathbb{R}^k$, the metric learning is to learn an appropriate distance metric between all data pairs for satisfying some distance constraints, such as the pair-wise distance ones. In general, given two sets of data pairs, the first one is the known similar pairs,
\begin{equation*}
  \mathcal{S} = \{(\bm{v}_i, \bm{v}_j) | \bm{v}_i \: \text{and} \: \bm{v}_j \: \text{are similar}\},
\end{equation*}
and the other one is the known dissimilar pairs,
\begin{equation*}
  \mathcal{D} = \{(\bm{v}_i, \bm{v}_j) | \bm{v}_i \: \text{and} \: \bm{v}_j \: \text{are dissimilar}\}.
\end{equation*}
Specifically, traditional approaches attempt to learn a Mahalanobis distance metric to make the distance in space smaller for similar pairs and larger for dissimilar pairs. The distance function is defined as,
\begin{equation*}
  d(\bm{v}_i, \bm{v}_j) = \sqrt{(\bm{v}_i - \bm{v}_j) \bm{M} (\bm{v}_i - \bm{v}_j)},
\end{equation*}
where $\bm{M}$ is a positive semi-definite matrix. A typical method is to globally solve the following convex optimization problem,
\begin{equation*}
\begin{aligned}
  \text{min}_{\bm{M}} \sum_{(\bm{v}_i, \bm{v}_i) \in \mathbb{S}} d(\bm{v}_i, \bm{v}_j)^2, \\
  s.t. \sum_{(\bm{v}_i, \bm{v}_j) \in \mathbb{D}} d(\bm{v}_i, \bm{v}_j)^2 \geq 1 \: \text{and} \: \bm{M} \succ 0,
\end{aligned}
\end{equation*}
where $\bm{M} \succ 0$ denotes matrix $\bm{M}$ is a positive semi-definite matrix.

\subsection{Factorization Machines}
Factorization Machines can work with any real valued feature vectors for prediction, which take  an \textbf{attribute vector} $\bm{x} \in \mathbb{R}^n$ as input and a real valued scalar as output. To intuitively illustrate the construction of $\bm{x}$, we take advantage of a concrete example from MovieLens. Suppose we would like to predict the rating for a movie in MovieLens. The selected attributes for prediction are user ID, movie ID and the leading actor ID, where the corresponding numbers are $N_u$, $N_m$ and $N_a$, respectively. Without loss of generality, we convert these three kinds of attributes into one-hot vectors and sequentially concatenate them to build a single attribute vector $\bm{x}$,
\begin{equation*}
\begin{aligned}
  \bm{x} = [ \, \underbrace{\cdot\cdot\cdot\, 00010}_{\text{length} N_u}, \, \underbrace{\cdot\cdot\cdot\, 00100}_{\text{length} N_m}, \, \underbrace{\cdot\cdot\cdot\, 01000}_{\text{length} N_a} \,] ,\\
  N_u + N_m + N_a = n,
\end{aligned}
\end{equation*}
where $i_u=1$, $i_m=2$ and $i_a=3$ represent the indices for the current sample of user ID, movie ID and the leading actor ID, respectively. FMs then estimate the rating by,
\begin{equation*}
  \hat{y}(\bm{x}) = w_0 + \sum_{i=1}^{n} w_i x_i + \sum_{i=1}^{n} \sum_{j=i+1}^{n}\langle \bm{v}_i, \bm{v}_j \rangle x_i x_j,
\end{equation*}
where $w_0$ denotes the global bias, $w_i$ models the strength of the $i$-th attribute $x_i$, and $\langle \bm{v}_i, \bm{v}_j \rangle$ models the second-order interactions between the $i$-th and $j$-th attributes. In the original FMs~\cite{fm}, $\bm{v}_i \in \mathbb{R}^k$ denotes the factorized feature vector for attribute $x_i$, and $\langle \bm{v}_i, \bm{v}_j \rangle$ represents the inner product of $\bm{v}_i$ and $\bm{v}_j$. To model more complex interactions between attributes, NFM~\cite{nfm} introduces deep learning into FMs and employs the fully connected layers to learn non-linear feature interactions,
\begin{equation*}
\begin{aligned}
  \hat{y}(\bm{x}) = w_0 + \sum_{i=1}^{n} w_i x_i + \bm{h}^T  \: \text{MLP}( \: f_{BI}(\bm{\mathbb{V}}_{\bm{x}})),   \\
  f_{BI}(\bm{\mathbb{V}}_{\bm{x}}) = \sum_{i=1}^{n} \sum_{j=i+1}^{n} x_i \bm{v}_i \odot x_j \bm{v}_j,
\end{aligned}
\end{equation*}
where MLP($\cdot$) and $\odot$ denote several fully connected deep layers and element-wise product, respectively. This approach is equal to inner product with non-linear transformations.
However, the inner product violates the triangle inequality and thus cannot well capture fine-grained attribute interactions~\cite{metriclearning}. With this observation, Pasricha et al.~\cite{transfm} recently proposed a TransFM method which employs the squared Euclidean distance function to replace the inner product for sequential recommendation,
\begin{equation*}
\begin{aligned}
  \hat{y}(\bm{x}) = w_0 + \sum_{i=1}^{n} w_i x_i + \sum_{i=1}^{n} \sum_{j=i+1}^{n} d(\bm{v}_i + \bm{v}_i^{\prime}, \bm{v}_j) x_i x_j, \\
  d(\bm{v}_i + \bm{v}_i^{\prime}, \bm{v}_j) = (\bm{v}_i + \bm{v}_i^{\prime} - \bm{v}_j)^T (\bm{v}_i + \bm{v}_i^{\prime} - \bm{v}_j),
\end{aligned}
\end{equation*}
where $\bm{v}_i$ and $\bm{v}_i^{\prime}$ are the embedding and translation feature vectors for attribute $x_i$, respectively.

\section{Proposed Method}\label{model}
Traditional FMs model the interactions between two attributes by the inner product, which does not satisfy the triangle inequality property of feature vectors. Recently, TransFM~\cite{transfm} has been proposed to use the squared Euclidean distance function to replace the inner product to model the interactions and achieved better performance. The reason is that the Euclidean distance obeys the triangle inequality property and thus can better capture the fine-grained relationships between feature vectors. However, the existing FMs only consider the interactions between attributes, while ignoring the ones among the features of each attribute (Remind that the attributes are  represented as latent feature vectors in FMs to estimate their interactions). We take a toy example for explanation. Suppose that we aim to predict the rating for a movie in MovieLens based on three attributes: \emph{user ID}, \emph{movie ID} and the \emph{leading actor ID}. In this case, we presume that the attribute \emph{leading actor ID} is parameterized by two kinds of features: \emph{acting performance} and \emph{actor popularity}. In existing FMs, they all model the interactions between the \emph{leading actor ID} and other attributes, i.e., inter-attribute interactions, while ignoring the correlations among the features of the \emph{leading actor ID} itself, which are also important for making predictions. In fact, for the attribute \emph{leading actor ID}, the correlations between \emph{acting performance} and \emph{actor popularity} feature, i.e., the intra-attribute interactions, are also essential to learn data-driven patterns effectively since the \emph{actor popularity} is usually positively related to the \emph{acting performance}.

In this section, we first introduce the overall framework of the proposed method, and then present two exemplar methods based on this framework to solve the above problems (e.g., linear and non-linear correlations between features). We then provide an optimization approach for the proposed method and the learning strategy adopted in this work. Finally, we theoretically prove that our method can be generalized to other distance functions as well as the vanilla FMs. The main notations involved in this paper are summarized in Table~\ref{tab:notation}.

\begin{table}
  \caption{Main notations involved in this paper.}
  \centering
  \scalebox{1.0}{
  \begin{tabular}{l|l}
    \hline
    Notations    & Definition and Description   \\
    \hline
    \hline
    $n$             & Length of the concatenated attribute vector \\
    $k$             & Dimension of the embedded features for an attribute \\
    $c$             & Scalar constant \\
    $\eta$          & Learning rate for gradient descent \\
    $w_0$           & Global bias in FMs \\
    $w_i$           & Weight of the $i$-th attribute ($i \geq 1$) \\
    $w_{ij}$        & Weight of the interaction of the $i$-th and $j$-th attribute \\
    \hline
    $\bm{x}$        & Vector of the concatenated attributes \\
    $\bm{v}_i$      & Vector of the $i$-th attribute \\
    $\bm{h}$        & Vector for computing the transformation weight \\
    $\bm{b}_l$      & Vector of the $l$-th layer's learnable bias \\
    $\bm{\theta}$   & Vector of learnable parameters \\
    \hline
    $\bm{L}$        & Matrix to parameterize the linear transformation \\
    $\bm{M}$        & Matrix for constraining the Mahalanobis distance \\
    $\bm{W}_l$      & Matrix of the $l$-th layer's learnable weights \\
    \hline
    $\mathbb{T}$    & Set of all the training instances \\
    $\mathcal{S}$   & Set of similar data pairs \\
    $\mathcal{D}$   & Set of dissimilar data pairs \\
    \hline
  \end{tabular}}
  \label{tab:notation}
\end{table}

\begin{figure*}
  \centering
  \includegraphics[width=\linewidth]{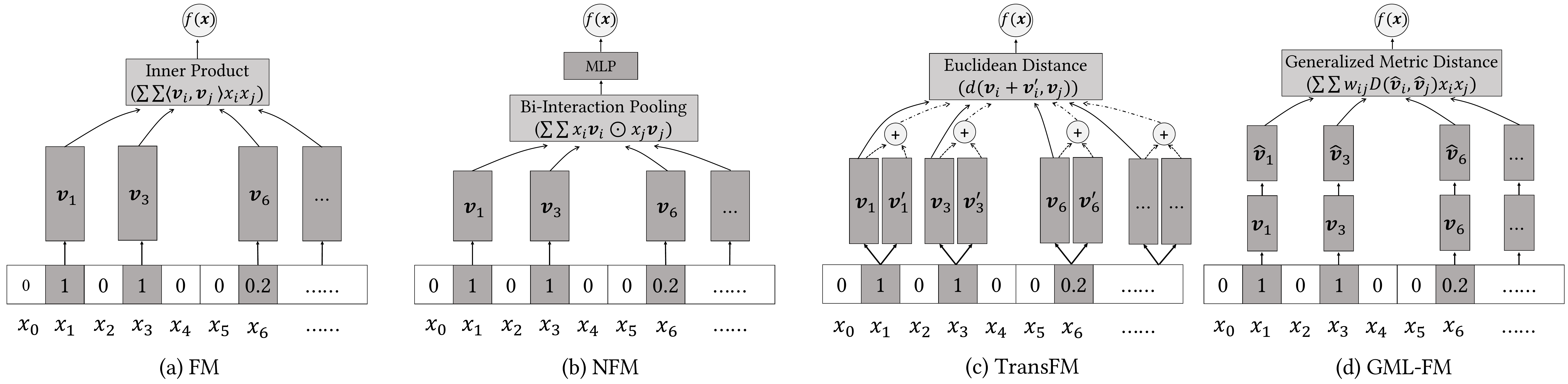}
  \caption{A visual comparison of FM, NFM, TransFM and GML-FM. (a) The FM uses Inner Product to model the second-order attribute interactions. (b) The NFM firstly designs an Bi-Interaction pooling layer to summarize the element-wise product between attribute embedding vectors, which is then input into a MLP framework. It can be seen as an extension of inner product. (c) TransFM takes the Euclidean Distance to compute the distance between ``the addition of the embedding vector and the translation vector from an attribute (i.e., $\bm{v}_i + \bm{v}_i^\prime$)'', and ``the embedding vector from another attribute (i.e., $\bm{v}_j$)''. (d) Different from the above these three models ignoring the feature-level interactions inside attributes, the proposed GML-FM capture this kind of interactions by transforming the original embedded vectors into a new space, where metric learning techniques can be effectively performed.}\label{fig:models}
\end{figure*}

\subsection{Model Formulation}
In order to model the feature correlations, we propose to use a generalized metric learning based approach to replacing the Euclidean distance function in FMs and dub this method as GML-FM. Similar to FMs, our proposed method could also take any real-value feature vectors as inputs. Formally, given an input vector $\bm{x} \in \mathbb{R}^n$, the target is estimated by,
\begin{equation}\label{equ:gmlfm}
  \hat{y}(\bm{x}) = w_0 + \sum_{i=1}^{n} w_i x_i + \sum_{i=1}^{n} \sum_{j=i+1}^{n} D(\bm{v}_i, \bm{v}_j) x_i x_j, \\
\end{equation}
where $\bm{v}_i$ and $\bm{v}_j$ are the factorized feature embeddings and $\bm{v}_i$,  $\bm{v}_j \in \mathbb{R}^k$, and $D(\cdot, \cdot)$ is the generalized distance function.

However, due to the inherent characteristic of distance functions, they are limited to non-negative values compared to the inner product, limiting the expressiveness of FMs. To solve this problem, we introduce a \emph{transformation weight} - $w_{ij}$, to convert the values of the second-order interactions to the real number values. Concretely, to avoid introducing more parameters and over-fitting, we leverage the existing embedded features $\bm{v}_i$ and $\bm{v}_j$, by combining them via the element-wise product, which is then converted to the transformation weight $w_{ij}$ by a trainable vector $\bm{h} \in \mathbb{R}^{k}$,
\begin{equation}\label{equ:gmlfm1}
  w_{ij} = \bm{h}^T (\bm{v}_i \odot \bm{v}_j). \\
\end{equation}
The transformation weight is to increase the representation ability of FM methods which is limited by the distance functions (i.e., distances are all non-negative real values). To this end, the second-order interactions of distance functions in FMs can achieve the same functionality of the ones based on the inner product. Therefore, the previous target function can be rewritten as,
\begin{equation}\label{equ:gmlfm2}
\begin{aligned}
  \hat{y}(\bm{x}) = w_0 + \sum_{i=1}^{n} w_i x_i + \sum_{i=1}^{n} \sum_{j=i+1}^{n} w_{ij} D(\bm{v}_i, \bm{v}_j) x_i x_j, \\
  w_{ij} = \bm{h}^T (\bm{v}_i \odot \bm{v}_j).
\end{aligned}
\end{equation}
Note that we do not explicitly define the similar and dissimilar attribute sets, where we leave the framework itself to automatically learn the correlation between attributes. In the next subsection, for the generalized distance function $D(\cdot, \cdot)$, we present two instances which can generalize the inner product as well as the Euclidean distance function.

\subsection{Generalized Metric Learning based FMs}
The Euclidean distance function is limited by its deficiency in modeling feature correlations. As these correlations play an important role in the target prediction, we thus devote our efforts to generalizing the Euclidean distance function to a generalized metric learning based one. In the following, we introduce two generalized metric learning based methods, which correspond to the linear and non-linear correlations between features, respectively.

\subsubsection{Mahalanobis Distance Function}
To effectively model the linear correlations between features, we adopt the Mahalanobis function~\cite{metriclearning} and form the distance function by,
\begin{equation}\label{equ:gml1}
\begin{aligned}
    D(\bm{v}_i, \bm{v}_j) = (\bm{v}_i - \bm{v}_j)^T \bm{M} (\bm{v}_i - \bm{v}_j),\\
    s.t., \bm{M} \succ 0,
\end{aligned}
\end{equation}
where $\bm{M} \in \mathbb{R}^{k \times k}$ is a transformation matrix. In particular, if $\bm{M}$ is a diagonal matrix, namely, the coordinate axis is orthogonal, the correlations between different features are independent. Nevertheless, different features may have positive or other linear correlations (recall the example that the positive correlation between the \emph{acting performance} and \emph{actor popularity} for the attribute of a movie's \emph{leading actor}), that is, the coordinate axis is not orthogonal. Therefore, it is sub-optimal to set $\bm{M}$ to be a diagonal matrix. Besides, the distance function should be non-negative which cannot be guaranteed by randomly initiating $\bm{M}$. Due to the aforementioned two considerations, $\bm{M}$ is set to be a positive semi-definite matrix~\cite{metriclearning}, which can be auto-learned from the training data. In the following, we show how to get the positive semi-definite matrix $\bm{M}$.

For those features with linear correlations, in order to correctly model the interactions, it is common to perform a linear transformation before the Euclidean distance,
\begin{equation}\label{equ:linear}
  \hat{\bm{v}}_i = \bm{L}\bm{v}_i,
\end{equation}
where $\bm{L} \in \mathbb{R}^{k \times k}$ and $\bm{L}$ parameterizes the transformation. Furthermore, the linear transformation can be expressed by,
\begin{equation}\label{equ:m}
  \bm{M} = \bm{L}^T \bm{L}.
\end{equation}
In this way, any matrix $\bm{M}$ from a real-valued matrix $\bm{L}$ is guaranteed to be positive semi-definite (i.e., to have no negative eigenvalues). This can be verified through the following proof.
\begin{proof}\renewcommand{\qedsymbol}{}
  For any real valued vector $\bm{x}$, the condition that $\bm{x}^T \bm{M} \bm{x} = \bm{x}^T \bm{L}^T \bm{L} \bm{x} = (\bm{L} \bm{x})^T (\bm{L} \bm{x}) \geq 0$ can always be satisfied. Therefore, matrix $\bm{M}$ is positive semi-definite.
\end{proof}

Notice that in this case, the Euclidean distance method is a special case of our approach by using the Mahalanobis distance. It can be obtained by simply setting the $\bm{M}$ in Equation~\ref{equ:gml1} as the identity matrix.

\subsubsection{DNN-based Distance Function}
The aforementioned method models the feature interactions in a linear way, which is insufficient for capturing the non-linear or other more complex correlations between features. For example, if we would like to recommend a piece of music to a user based on its \emph{elements} attribute, which can be represented by the \emph{rhythm} and \emph{melody}, it may not be optimal to combine these two features independently or linearly, since different types of rhythms can co-exist with the same melody and vice versa. To capture such complex correlations, it is better to model the feature interactions in a non-linear way. For modeling the interactions and obtaining better fusion features, we refer to DNNs, which apply multi-layer of non-linear interactions to model feature interactions and have been shown to be very effective~\cite{ncf, alstp, conversational}. Specifically, original $\bm{v}_i$ is transformed by a deep neural network,
\begin{equation}\label{equ:gml22}
  \hat{\bm{v}}_i = \sigma_L (\bm{W}_L(... \sigma_1 (\bm{W}_1 \bm{v}_i + \bm{b}_1)) + \bm{b}_L),
\end{equation}
where $\bm{W}_l$ and $\bm{b}_l$ denote the weight matrix and bias vector for the $l$-th fully connected layer, respectively. And $L$ is the number of deep layers used in the network. A dropout layer~\cite{dropout} is deployed between each contiguous fully connected layers to prevent over-fitting. For simplicity, all the learnable weight and bias are set to be the same size in this work, namely, $\bm{W}_l \in \mathbb{R}^{k\times k}$ and $\bm{b}_l \in \mathbb{R}^k$. And $\sigma_l(\cdot)$ is the activation function, which could be sigmoid, hyperbolic tangent (tanh), rectified linear unit (ReLU), etc. In this work, we take the tanh as the activation function for all layers, which can map the input in the range of -1 to 1. With this deep non-linear neural network, it is expected that more complex correlations between features inside attributes can be captured.

After this procedure, the distance function becomes,
\begin{equation}\label{equ:gml2}
    D(\bm{v}_i, \bm{v}_j) = (\hat{\bm{v}}_i - \hat{\bm{v}}_j)^T(\hat{\bm{v}}_i - \hat{\bm{v}}_j),
\end{equation}
where both $\hat{\bm{v}}_i$ and $\hat{\bm{v}}_j$ are the learned features via non-linear transformations from $\bm{v}_i$ and $\bm{v}_j$, respectively.

In particular, the Euclidean distance can be easily recovered by setting all the weight matrices $\bm{W}_l$ in Equation~\ref{equ:gml22} to be identity matrix, the bias vectors $\bm{b}_l$ to be zeros and the activation functions $\sigma_l(\cdot)$ to be the identity function.

A visual comparison of our method and three state-of-the-art FM models is shown in Figure~\ref{fig:models}. As can be observed, all the three previous researches (i.e., FM, NFM and TransFM) focus on the interactions on the attribute level, i.e., inter-attribute interaction. In contrast, our proposed GML-FM method takes the feature-level interactions inside attributes into consideration, i.e., intra-attribute interaction. We empirically demonstrate that the feature-level interactions are important for better modeling the complex and rich interactions of real-world data in Section~\ref{results}.

\subsection{Proposed Efficient Solution}
Formally, if the proposed method is computed in a straight way, the time cost will be in linear time $O((kn)^2)$, where $k$ is the embedding size and $n$ is the length of the concatenated attribute vector, which is too expensive. In the following, we will provide an effective approach to efficiently simplifying the proposed method equations. We theoretically analyze that our proposed solution can greatly reduce the time complexity of the proposed method.

In the next, we first present a simplified general form of the second-order interaction and then provide the proposed approaches for both the Mahalanobis and DNN distance functions, which can significantly simplify the computation and reduce the time complexity. The general form of the second-order interaction of our proposed method is given in Equation~\ref{equ:simfm_overall},
\begin{equation}\label{equ:simfm_overall}
\begin{aligned}
  f(\bm{x}) &= \sum_{i=1}^{n} \sum_{j=i+1}^{n} \bm{h}^{T} (\bm{v}_i \odot \bm{v}_j) D(\bm{v}_i, \bm{v}_j) x_i x_j, \\
            &= \frac{1}{2} (\sum_{i=1}^{n} \sum_{j=1}^{n} \bm{h}^{T} (\bm{v}_i \odot \bm{v}_j) D(\bm{v}_i, \bm{v}_j) x_i x_j - \\
            & \: \: \: \: \: \: \sum_{i=1}^{n} \bm{h}^{T} (\bm{v}_i \odot \bm{v}_i) D(\bm{v}_i, \bm{v}_i) x_i x_i), \\
            &= \frac{1}{2} \sum_{i=1}^{n} \sum_{j=1}^{n} \bm{h}^{T} (\bm{v}_i \odot \bm{v}_j) D(\bm{v}_i, \bm{v}_j) x_i x_j. \\
\end{aligned}
\end{equation}
Note that $D(\cdot, \cdot)$ represents the distance function, which will be zero if the two inputs are same (i.e., $D(\bm{v}_i, \bm{v}_i)=0$). In the following, we illustrate the two proposed generalized metric learning based FMs and the corresponding simplified form.

\subsubsection{Mahalanobis Distance Function} We show the derivation of the Mahalanobis distance function with the transformation weight in Equation~\ref{equ:simfm_mdis1}. With the simplification of Equation~\ref{equ:simfm_overall}, the second-order interaction of the original Mahalanobis distance function based FMs can be rewritten as (please refer to Appendix~\ref{appendix} for detailed derivation),

\begin{equation}\label{equ:simfm_mdis1}
\begin{aligned}
  f(\bm{x}) &= \sum_{i=1}^{n} \sum_{j=i+1}^{n} \bm{h}^{T} (\bm{v}_i \odot \bm{v}_j) D(\bm{v}_i, \bm{v}_j) x_i x_j, \\
            &= \sum_{j=1}^{n} \bm{v}_j^T x_j \sum_{i=1}^{n} diag(\bm{h}) \bm{v}_i \bm{v}_i^T \bm{M} \bm{v}_i x_i - \\
            & \: \: \: \: \:   \sum_{j=1}^{n} \bm{v}_j^T diag(\bm{h}) ( \sum_{i=1}^{n} \bm{v}_i \bm{v}_i^T x_i) \bm{M} \bm{v}_j x_j, \\
\end{aligned}
\end{equation}
where $diag(\bm{h})$ is an operation to convert the vector $\bm{h}$ to a diagonal matrix. Based on this simplification, the two summations of $i$ and $j$ can be computed independently or sequentially, where the original two summations are performed in a nested structure, which is the major time-consuming factor for the $O((kn)^2)$ computation in Equation~\ref{equ:simfm_overall}. In the next, we will elaborate how this solution can achieve the $O(k^2n)$ time complexity.

Notice that the time complexity of the two elements on the right hand side (RHS) of Equation~\ref{equ:simfm_mdis1} is equal. Therefore, we only analyze the time complexity of the first one on the RHS as an example in the next. This element is a computation of two sums and the main cost is from the second one (i.e., $\sum_{i=1}^{n} diag(\bm{h}) \bm{v}_i \bm{v}_i^T \bm{M} \bm{v}_i x_i$). And the computation of the second one is composed of 3 vector element-wise product or inner product and one vector-matrix multiplication. The time complexity is therefore $O((3k + k^2)n) = O(k^2n)$. Recall that the original time complexity of the GML-FM is $O(k^2n^2)$. Since $k$ is usually much smaller than that of $n$ ($k$ is usually of tens or hundreds. In contrast, $n$ is often tens of thousands, or even tens of millions), we argue that the proposed solution can significantly reduce the original time complexity of the proposed method.

\subsubsection{Generalized Metric Distance with Deep Neural Networks}
It is well-known that training DNNs is very expensive compared to traditional shallow models. We thus propose to simplify the original DNN-based model equations. Similar as the previous one, the detailed derivation under this setting is,
\begin{equation}\label{equ:simfm_mdis2}
\begin{aligned}
  f(\bm{x}) &= \sum_{i=1}^{n} \sum_{j=i+1}^{n} \bm{h}^{T} (\bm{v}_i \odot \bm{v}_j) D(\bm{v}_i, \bm{v}_j) x_i x_j, \\
            &= \sum_{i=1}^{n} \sum_{j=i+1}^{n} \bm{h}^{T} (\bm{v}_i \odot \bm{v}_j) (\hat{\bm{v}}_i - \hat{\bm{v}}_j)^T(\hat{\bm{v}}_i - \hat{\bm{v}}_j) x_i x_j, \\
            &= \frac{1}{2} \sum_{i=1}^{n} \sum_{j=1}^{n} \bm{v}_i^T diag(\bm{h}) \bm{v}_j (\hat{\bm{v}}_i^T \hat{\bm{v}}_i - 2\hat{\bm{v}}_i^T \hat{\bm{v}}_j + \hat{\bm{v}}_j^T \hat{\bm{v}}_j) x_i x_j,  \\
            &= \sum_{j=1}^{n} \bm{v}_j^T x_j \sum_{i=1}^{n} diag(\bm{h}) \bm{v}_i \hat{\bm{v}}_i^T \hat{\bm{v}}_i x_i -  \\
            & \: \: \: \: \: \: \sum_{j=1}^{n} \bm{v}_j^T diag(\bm{h}) ( \sum_{i=1}^{n} \bm{v}_i \hat{\bm{v}}_i^T x_i) \hat{\bm{v}}_j x_j, \\
\end{aligned}
\end{equation}
where $\hat{\bm{v}}_i$ and $\hat{\bm{v}}_j$ are produced by deep neural networks,
\begin{equation}
  \hat{\bm{v}}_i = \sigma_L (\bm{W}_L(... \sigma_1 (\bm{W}_1 \bm{v}_i + \bm{b}_1)) + \bm{b}_L).
\end{equation}
Within the deep neural networks, the matrix-vector multiplication of weight matrices and input features is the main operation which can be computed in $O(k^2)$ (we set all the weight matrices to be in $\mathbb{R}^{k \times k}$). In short, the same as above, the overall time complexity for evaluating the GML-FM method is $O(k^2n)$.

\subsection{Learning Strategy}
GML-FM can be applied to a variety of prediction tasks, including classification, regression and ranking. In this work, we adopt a commonly used regression objective function (i.e., the squared loss),
\begin{equation}\label{equ:loss}
  L_{reg} = \sum_{\bm{x} \in \mathbb{T}} {(\hat{y}(\bm{x}) - y(\bm{x}))^2},
\end{equation}
where $\mathbb{T}$ represents all the training instances and $y(\bm{x})$ denotes the corresponding target of $\bm{x}$. To optimize the objective, we employ the stochastic gradient descent (SGD), a universal solver for optimizing machine learning models. The SGD updates the model parameters towards the direction of the negative gradients. Normally, a mini-batch of training instances is selected for training and optimizing model parameters,
\begin{equation}\label{equ:sgd}
  \bm{\theta} = \bm{\theta} - \eta \cdot 2(\hat{y}(\bm{x}) - y(\bm{x})) \frac{d\hat{y}(\bm{x})}{d \bm{\theta}},
\end{equation}
where $\bm{\theta}$ denotes the learnable parameters (e.g., $\bm{W}$ in deep neural networks), and $\eta$ is the learning rate deciding the step size of gradient descent.

\subsection{Generalization to More Distance Functions}
In addition to modeling the correlations between features, our method can also cover other common distance functions in the vector space to learn the inter-attribute interactions. Specifically, we employ the Minkowski distance function to illustrate the different variants,
\begin{equation*}
    D(\bm{v}_i, \bm{v}_j) = (\sum_{m=1}^k |\hat{v}_i^m - \hat{v}_j^m|^p)^{\frac{1}{p}},
\end{equation*}
where $\hat{v}_i^m$ denotes the $m$-th element in the transformed features $\hat{\bm{v}}_i$. In this way, the proposed method can express the following distance functions:
\begin{itemize}
    \item Manhattan distance when $p=1$;
    \item Euclidean distance when $p=2$;
    \item Chebyshev distance when $p=\infty$.
\end{itemize}
Moreover, if we reduce to the inner product fashioned inter-attribute interaction modeling, the proposed method can also cover the Cosine distance function as follows,
\begin{equation*}
    D(\bm{v}_i, \bm{v}_j) = \frac{\hat{\bm{v}}_i^T \hat{\bm{v}}_j}{||\hat{\bm{v}}_i||_2 ||\hat{\bm{v}}_i||_2}.
\end{equation*}

\subsection{Relation with FMs}
In this subsection, we prove that the vanilla FMs are a special case of our GML-FM method by setting $w_{ij}$ = 1 and $D(\cdot, \cdot)$ to be the Euclidean distance. By expanding the squared function and constraining all the $\| \bm{v}_i \|^2$ to be a constant, e.g., 1 (which gives more geometry meaning since the vector $\bm{v}_i$ is an orthogonal basis),  we can derive that,
\begin{equation}\label{equ:generalfm}
\begin{aligned}
  \hat{y}(\bm{x})   &= w_0 + \sum_{i=1}^{n} w_i x_i + \sum_{i=1}^{n} \sum_{j=i+1}^{n} w_{ij} D(\bm{v}_i, \bm{v}_j) x_i x_j, \\
                    &= w_0 + \sum_{i=1}^{n} w_i x_i + \sum_{i=1}^{n} \sum_{j=i+1}^{n} (\bm{v}_i - \bm{v}_j)^2 x_i x_j, \\
                    &= w_0 + \sum_{i=1}^{n} w_i x_i + \sum_{i=1}^{n} \sum_{j=i+1}^{n} (\| \bm{v}_i \|^2 + \| \bm{v}_j \|^2 - \\
                    & \: \: \: \: \: \: 2 \langle \bm{v}_i, \bm{v}_j \rangle) x_i x_j, \\
                    &= w_0 + \sum_{i=1}^{n} w_i x_i + c_1 \sum_{i=1}^{n} \sum_{j=i+1}^{n} \langle \bm{v}_i, \bm{v}_j \rangle x_i x_j + c_2,
\end{aligned}
\end{equation}
where $c_1$ and $c_2$ are constants. It is worth pointing out that, to the best of our knowledge, this is the first work proving that metric learning based FMs can generalize the vanilla FMs. 
\section{Experimental Setup}\label{setup}
We evaluate the effectiveness of our method on two common recommendation tasks: rating prediction and top-n recommendation. The former attempts to predict the rating that a user would give to an unseen item, and the latter one is to rank unseen items to a target user based on his/her preference. For the latter, our method rank the items according to the predicted ratings.

\begin{table}
  \caption{Statistics of the evaluation datasets. The first three datasets are from the Amazon dataset, followed by the next two from the Mercari dataset. \#feature-dim represents the number of dimension of the concatenated attribute vector.}
  \centering
  \scalebox{0.9}{
  \begin{tabular}{|l|r|r|r|r|r|}
    \hline
    Datasets    &\#users&\#items    &\#attribute-dim     &\#instances   &sparsity    \\
    \hline
    \hline
    Auto                        & 2,928     & 1,835     & 5,220         & 20,473       & 99.62\%      \\
    Office                      & 4,905     & 2,420     & 7,620         & 53,258       & 99.55\%      \\
    Clothing                    & 39,387    & 23,033    & 64,473        & 278,677      & 99.96\%      \\
    Ticket                      & 3,855     & 45,998    & 49,977        & 46,712       & 99.97\%      \\
    Books                       & 26,080    & 367,968   & 394,177       & 373,790      & 99.99\%      \\
    MovieLens                   & 6,040     & 3,706     & 10,070        & 1,000,209    & 95.53\%      \\
    \hline
  \end{tabular}}
  \label{tab:dataset}
\end{table}

\subsection{Datasets}
The experiments are conducted on three datasets, including two publicly available datasets (i.e., the Amazon Product dataset~\cite{amazon2} and the MovieLens dataset~\cite{movielens}) and a newly collected dataset which is used to further explore the effectiveness of the proposed method under a very sparse setting. Details of these three datasets are as follows:

\textbf{Amazon.} The Amazon dataset\footnote{http://jmcauley.ucsd.edu/data/amazon/.} contains product reviews, ratings and metadata, with the user interactions spanning from May 1996 to July 2014. In our experiments, we adopted the 5-core version datasets. And three categories are used in our experiments: Auto, Office and Clothing. For attribute extraction, we leveraged the user ID, item ID and item sub-category as the experimented attributes.

\textbf{MovieLens.} The MovieLens dataset\footnote{https://grouplens.org/datasets/movielens/.} has been long recognized as a protocol dataset for evaluating recommendation algorithms. We used one of the most stable dataset - MovieLens 1M, which contains rich side information, including user gender, user age, user occupation, and item genres.

\textbf{Mercari.} This dataset is collected from a second-hand product trading platform - Mercari\footnote{https://mercari.com/.}, including 265.4 million items, 10.7 million buyers, 8.4 million sellers, and user behaviors spanning from Nov. 2016 to Oct. 2018. This dataset includes user information (ID, status), product metadata (product ID, seller ID, price, brand, category, condition, size, description, status), product shipping information (methods, origin, duration, payer), and user behaviors (liking, listing, purchasing, and click). Two categories are adopted: Ticket and Books. And we utilized the purchase interactions and kept the users with at least five items in their purchase history. For side information, we selected a set of item features, including category, conditions, shipping method, shipping origin and shipping duration.

The basic statistics are summarized in Table~\ref{tab:dataset}.

\subsection{Compared Baselines}
We compared the proposed method with two groups of recommendation methods: MF-based and FM-based. The former one contains MF, PMF, NCF, NGCF and BPR-MF, which considers the user-item interactions only, without any side information. And the latter one includes the state-of-the-art FM methods - FM, DeepFM, xDeepFM, NFM, AFM and TransFM, which leverages the rich side information for recommendation. Specifically, the MF and PMF are adopted to evaluate the rating prediction. The NCF, NGCF and BPR are designed for evaluating the top-n recommendation, and the other six baselines are exploited for both tasks.

\textbf{Unique Baselines for Rating Prediction}:

\textbf{MF}~\cite{mf} factorizes the rating matrix into latent vectors of users and items, and uses the inner product between users and items to estimate the interactions.

\textbf{PMF}~\cite{pmf} extends MF methods by adopting a probabilistic linear model with Gaussian observation noise~\cite{prob} for user and item feature vectors.

\textbf{Unique Baselines for Top-n Recommendation}:

\textbf{NCF}~\cite{ncf} extends MF methods with deep neural networks, where the inner product between users and items is replaced with non-linear interactions. NCF is devised in a point-wise learning to rank fashion.

\textbf{NGCF}~\cite{ngcf} proposes to integrate the user-item interactions - the bipartite graph structure - into the embedding process. It aims to expressively model the higher-order connectivity in the user-item graph.

\textbf{BPR-MF}~\cite{bpr} leverages the Bayesian Personalized Ranking (BPR) framework with MF as the underlying model. It adopts the pair-wise learning to rank strategy.

\textbf{Common Baselines for Both Tasks}:

\textbf{FM.} This is the standard FM method, which is originally proposed for recommendation~\cite{fm}. In experiments, we used the official implementation LibFM\footnote{http://libfm.org/.} and adopted the SGD optimizer in accordance with other methods.

\textbf{DeepFM}~\cite{deepfm} combines FMs for recommendation and deep learning for feature learning in a Wide \& Deep~\cite{wide} architecture. It exploits both the low- and high-order attribute interactions.

\textbf{xDeepFM}~\cite{xdeepfm} replaces the FM part in DeepFM with a novel Compressed Interaction Network (CIN). The CIN is designed to generate the attribute interactions in an explicit fashion and at the vector-wise level.

\textbf{NFM}~\cite{nfm} extends the inner product interaction in FMs with non-linear multi-layer perceptron. This work brings together the effectiveness of linear factorization machines with the strong representation ability of non-linear neural networks.

\textbf{AFM}~\cite{afm} automatically learns the importance of each attribute interaction via deep neural networks (i.e., attention networks).

\textbf{TransFM}~\cite{transfm} is a recently proposed metric learning based FM. It replaces the inner product in FM with the Euclidean distance function. We adapted it from the sequential recommendation to the general recommendation setting, where we removed the constriction that two items have to be sequentially adjacent.

\subsection{Evaluation Protocols}
As the objective of rating prediction and top-n recommendation is different, we therefore used separate settings for these two tasks.
\subsubsection{Rating Prediction}
For rating prediction task, as the dataset contains positive instances only, we thus randomly sampled two negative instances for each positive instance (the user-item pairs which are not interacted by the current user) to ensure the generalization of the predictive model. We set the positive instance score with 1 and negative with -1 for implicit feedback setting~\cite{implicit}. Moreover, the dataset is randomly split into 70\% training, 20\% validation and 10\% testing, where the validation set is used for tuning the hyper-parameters and the final results are reported on the testing set.

The evaluation metric we adopted is the commonly used Root Mean Square Error (\textbf{RMSE}), where a lower score denotes the better model performance.
\subsubsection{Top-n Recommendation}
To evaluate the model performance, we followed the widely used \emph{leave-one-out} evaluation~\cite{bpr,ncf}, where the latest interaction data of each user is used for testing and all the previous interactions are used for training. As all the datasets contain positive interactions only, we randomly sampled two negative instances to pair with one positive instance in the training set to ensure the generalization of the models~\cite{nfm,afm}. For fair comparison, we leveraged the same positive and negative instances for all the models.

We applied two standard metrics in evaluation: Hit Ratio (\textbf{HR}) and Normalized Discounted Cumulative Gain (\textbf{NDCG})~\cite{ncf}, where the first indicates the percentage of items which are recommended correctly, while the latter considers the position of positive items in the ranking list. Since it is too time-consuming to rank all the items for each user during testing, we followed the common strategy~\cite{ncf,rank} to randomly select 99 items that are not purchased by the candidate user, and truncated the top 10 ranked items for both metrics. We calculated the two metrics for each user and reported the average scores.

\subsection{Parameter Settings}
We initialized all the parameters with normal distribution (with a mean of 0 and standard deviation of 0.01) and optimized the whole model with the Adam optimization~\cite{adam}. The learning rate is tuned in the range of [0.0001, 0.001, 0.01, 0.1], and dropout is [0, 1.0] with a step size of 0.1. The batch size is fixed to be 256. We carefully tuned the number of deep layers from 0 to 3 and the embedding size from 4 to 512. The code has been released for the reproductivity of this work.

\section{Experimental Results}\label{results}

\begin{table*}
  \caption{Overall performance comparison over six datasets on the rating prediction task. Symbol $\dag$ and $\ast$ denote the statistical significance with two-sides t-test of $p<0.01$, $p<0.05$, respectively, compared to the best baseline. The best performance is highlighted in boldface.}\label{tab:baseline2}
  \centering
  \scalebox{1.0}{
  \begin{tabular}{|l|c|c|c|c|c|c|}
    \hline
                            & MovieLens         & Amazon-Office     & Amazon-Clothing   & Amazon-Auto   & Mercari-Ticket    & Mercari-Books  \\
    \hline
    Model                   & RMSE              & RMSE              & RMSE              & RMSE          & RMSE              & RMSE            \\
    \hline
    \hline
    MF                      & 0.6389            & 0.8415            & 0.9619            & 0.9762        & 0.9974            & 0.9987           \\
    PMF                     & 0.6456            & 0.8380            & 0.9417            & 0.9468        & 0.9895            & 0.9993           \\
    \hline
    LibFM                   & 0.6592            & 0.8686            & 0.9213            & 0.9369        & 0.9731            & 0.9688            \\
    NFM                     & \textbf{0.6377}   & 0.8584            & 0.9147            & 0.9136        & 0.9218            & 0.8847            \\
    AFM                     & 0.6780            & 0.8663            & 0.9212            & 0.9315        & 0.7915            & 0.8260             \\
    TransFM                 & 0.6617            & 0.8616            & 0.9155            & 0.9282        & 0.9725            & 0.9697             \\
    DeepFM                  & 0.6402            & 0.8179            & 0.8940            & 0.9161        & 0.9444            & 0.7650             \\
    xDeepFM                 & 0.6412            & 0.8214            & 0.8961            & 0.9126        & 0.9372            & \textbf{0.7272}             \\
    \hline
    GML-FM$_{md}$           & 0.6472            & 0.8319            & 0.8930            & 0.9050        & 0.7655            & 0.7902             \\
    GML-FM$_{dnn}$          & 0.6446                          & \textbf{0.8153}$^\ast$             & \textbf{0.8861}$^\dag$
                            & \textbf{0.8822}$^\dag$                & \textbf{0.7572}$^\dag$            & 0.7892$^\dag$                  \\
    \hline
  \end{tabular}}
\end{table*}

\begin{table*}
  \caption{Overall performance comparison over six datasets on the top-n recommendation task. Symbol $\dag$ and $\ast$ denote the statistical significance with two-sides t-test of $p<0.01$, $p<0.05$, respectively, compared to the best baseline. The best performance is highlighted in boldface.}\label{tab:baseline}
  \centering
  \scalebox{1.0}{
  \begin{tabular}{|l|cc|cc|cc|cc|cc|cc|}
    \hline
                            & \multicolumn{2}{c|}{MovieLens}          & \multicolumn{2}{c|}{Amazon-Office}  & \multicolumn{2}{c|}{Amazon-Clothing}
                            & \multicolumn{2}{c|}{Amazon-Auto}        & \multicolumn{2}{c|}{Mercari-Ticket} & \multicolumn{2}{c|}{Mercari-Books}    \\
    \hline
    Model                   & HR     & NDCG   & HR     & NDCG   & HR     & NDCG   & HR     & NDCG   & HR     & NDCG   & HR     & NDCG               \\
    \hline
    \hline
    NCF                     & 0.5644 & 0.2898          & 0.2532 & 0.1215 & 0.2737 & 0.1496 & 0.2538 & 0.1329          & 0.3074 & 0.1588 & 0.4274 & 0.2448             \\
    BPR-MF                  & 0.6573 & 0.3814          & 0.2612 & 0.1300 & 0.2743 & 0.1710 & 0.3740 & \textbf{0.2264} & 0.1222 & 0.0603 & 0.1289 & 0.0759             \\
    NGCF                    & 0.5503 & 0.2799          & 0.2609 & 0.1278 & 0.3012 & 0.1746 & 0.3221 & 0.1786          & 0.1010 & 0.0409 & 0.3409 & 0.1717             \\
    \hline
    LibFM                   & 0.3538 & 0.1800          & 0.2100 & 0.0980 & 0.2912 & 0.1621 & 0.3026 & 0.1662          & 0.1320 & 0.0622 & 0.1080 & 0.0489             \\
    NFM                     & 0.6701 & \textbf{0.3896} & 0.2599 & 0.1199 & 0.2766 & 0.1517 & 0.3029 & 0.1683          & 0.1863 & 0.0865 & 0.1711 & 0.0770             \\
    AFM                     & 0.6182 & 0.3307          & 0.2540 & 0.1240 & 0.2968 & 0.1689 & 0.2811 & 0.1465          & 0.4169 & 0.2149 & 0.3328 & 0.1601             \\
    TransFM                 & 0.6584 & 0.3779          & 0.2722 & 0.1338 & 0.3413 & 0.1897 & 0.3173 & 0.1734          & 0.2285 & 0.1303 & 0.2514 & 0.1727             \\
    DeepFM                  & 0.6650 & 0.3792          & 0.3062 & 0.1567 & 0.3086 & 0.1680 & 0.3272 & 0.1735          & 0.4088 & 0.1798 & 0.4666 & 0.2433             \\
    xDeepFM                 & 0.6609 & 0.3813          & 0.3031 & 0.1539 & 0.3221 & 0.1709 & 0.3300 & 0.1823          & 0.4030 & 0.1809 & \textbf{0.5337} & \textbf{0.2897}             \\
    \hline
    GML-FM$_{md}$           & 0.6608 & 0.3742          & 0.3038 & 0.1537 & 0.3465 & 0.1984 & 0.3463 & 0.1993          & 0.5349 & 0.2478 & 0.4324 & 0.2086             \\
    GML-FM$_{dnn}$         & \textbf{0.6709}          & 0.3889 & \textbf{0.3354}$^\dag$ & \textbf{0.1756}$^\dag$     & \textbf{0.3794}$^\dag$ & \textbf{0.2160}$^\dag$
                            & \textbf{0.4133}$^\ast$   & 0.2177 & \textbf{0.5782}$^\dag$ & \textbf{0.2894}$^\dag$     & 0.4458$^\dag$ & 0.2143$^\dag$             \\
    \hline
  \end{tabular}}
\end{table*}

In this section, we report and analyze the experimental results. Particularly, we focus on the following research questions:
\begin{itemize}
  \item \textbf{RQ1}: Can our model outperform the state-of-the-art recommendation baselines on both tasks?
  \item \textbf{RQ2}: How different distance functions and different layers in the DNN-based distance function affect the model performance?
  \item \textbf{RQ3}: How does the embedding size affect both the proposed model and baselines' performance?
  \item \textbf{RQ4}: How does the proposed method perform with different attributes?
  \item \textbf{RQ5}: Can the proposed method show advantage under cold-start settings?
  \item \textbf{RQ6}: What do the proposed metric learning based and the previous FM methods learn in the embedded latent space?
\end{itemize}
\subsection{Performance Comparison (RQ1)}\label{res:all}
Table~\ref{tab:baseline2} and Table~\ref{tab:baseline} show the performance of our methods and the baselines across all the datasets on the rating prediction and top-n recommendation tasks, respectively. Note that the GML-FM$_{md}$ and GML-FM$_{dnn}$ represent our method with Mahalanobis distance and DNN-based distance funtions, respectively. In addition, we also conducted pairwise significance test between our method and the baseline with the best performance. Note that a smaller RMSE in Table~\ref{tab:baseline2} and a larger HR or NDCG in Table~\ref{tab:baseline} denote the better performance. The key observations are as follows.

Firstly, our proposed methods, especially GML-FM$_{dnn}$, outperform all the baselines across five datasets consistently and significantly (slightly worse on the MovieLens dataset on the rating prediction task) except for the Mercari-Books dataset. In particular, the sparser the dataset is, the larger improvement our method can achieve. Specifically, the sparsity of the three datasets (i.e., MovieLens, Amazon-Office, Mercari-Ticket) is 95.53\%, 99.55\% and 99.97\%, respectively, and the corresponding improvements (absolute) of HR on the top-n recommendation task are 0.08\%, 2.92\% and 16.13\% compared to the best baseline, respectively. This demonstrates the advantage of our method on sparse datasets.

Secondly, our proposed GML-FM$_{md}$ method can also outperform all the baselines on most occasions. However, the performance of GML-FM$_{md}$ is inferior than that of GML-FM$_{dnn}$, which proves that the feature correlations in real-world data are often non-linear and quite complex.

Finally, the FM-based baselines surpass the MF-based ones (i.e., MF and PMF for rating prediction, NCF, NGCF and BPR for top-n recommendation) on four sparser datasets (i.e., Amazon-Clothing, Amazon-Auto, Mercari-Ticket and Mercari-Books). It is expected because FM-based methods exploit more side information, which can improve the recommendation accuracy, especially for sparse datasets. This point has been widely proved in studies~\cite{nfm,afm}. It is also worth noting that although the HR of BPR-MF is worse than other methods on top-n recommendation task, its NDCG is very competitive except for the two sparsest datasets. It is because NDCG takes the position of positive items into consideration, and pairwise learning methods are more suitable for the ranking task.

\subsection{Ablation Study (RQ2)}\label{res:ablation}
To validate the effectiveness of the Mahalanobis distance and different layers of deep neural networks, we justified the variants of our proposed method and reported the final performance on two datasets (i.e., MovieLens and Mercari-Ticket) of the two tasks in Table~\ref{tab:ablation}. The main observations are four-fold:

\begin{table*}
  \caption{Comparison among the variants of our proposed method. Note that `M' denotes the Mahalanobis matrix, the model without the introduced transformation weight and Mahalanobis matrix is reduced to the conventional Euclidean distance function without considering the inter-attribute interactions. And the model with \#layers equal to 0 is equivalent to the Euclidean distance function with the transformation weight. Other distance functions we tested are under the \#layers $==1$ setting. }\label{tab:ablation}
  \centering
  \scalebox{1.0}{
    \begin{tabular}{|c|c|c|c|cc|cc|}
    \hline
    \multicolumn{2}{|c|}{}                  & \multicolumn{2}{c|}{Rating Prediction}    &\multicolumn{4}{c|}{Top-n Recommendation} \\
    \hline
    \multicolumn{2}{|c|}{}                  & MovieLens & Mercari-Ticket                & \multicolumn{2}{c|}{MovieLens}    & \multicolumn{2}{c|}{Mercari-Ticket}\\
    \hline
    \multicolumn{2}{|c|}{Models}            & RMSE      & RMSE                          & HR        & NDCG                  & HR            & NDCG              \\
    \hline
    \hline
    \multicolumn{2}{|c|}{w/o. weight \& M}  & 0.6861    & 1.0693                        & 0.6435    & 0.3702                & 0.1699        & 0.0743            \\
    \multicolumn{2}{|c|}{w/. M only}        & 0.6815    & 0.9627                        & 0.6091    & 0.3446                & 0.0423        & 0.0181            \\
    \multicolumn{2}{|c|}{w/. weight \& M}   & 0.6469    & 0.7736                        & 0.6608    & 0.3742                & 0.5349        & 0.2478            \\
    \hline
    \multirow{4}{*}{\#layers}   & 0         & 0.6475    & 0.7832                        & 0.6553    & 0.3762                & 0.5245        & 0.2444            \\
                                & 1         & 0.6446    & 0.7579                        & 0.6709    & 0.3889                & 0.5782        & 0.2894            \\
                                & 2         & 0.6478    & 0.7456                        & 0.6732    & 0.3879                & 0.5857        & 0.2963            \\
                                & 3         & 0.6492    & 0.7545                        & 0.6695    & 0.3853                & 0.5562        & 0.2691            \\
    \hline
    \multirow{4}{*}{distance functions} & Manhattan        & 0.6832    & 0.7903                        & 0.6498    & 0.3799                & 0.5335        & 0.2701            \\
    & Euclidean         & 0.6446    & 0.7579                        & 0.6709    & 0.3889                & 0.5782        & 0.2894            \\
    & Chebyshev          & 0.7112    & 0.7943                        & 0.6406    & 0.3731                & 0.5134        & 0.2567            \\
    & Cosine          & 0.7018    & 0.7965                        & 0.6330    & 0.3725                & 0.5053        & 0.2509            \\
    \hline
    \end{tabular}}
\end{table*}

\begin{figure*}
  \centering
  \includegraphics[width=\linewidth]{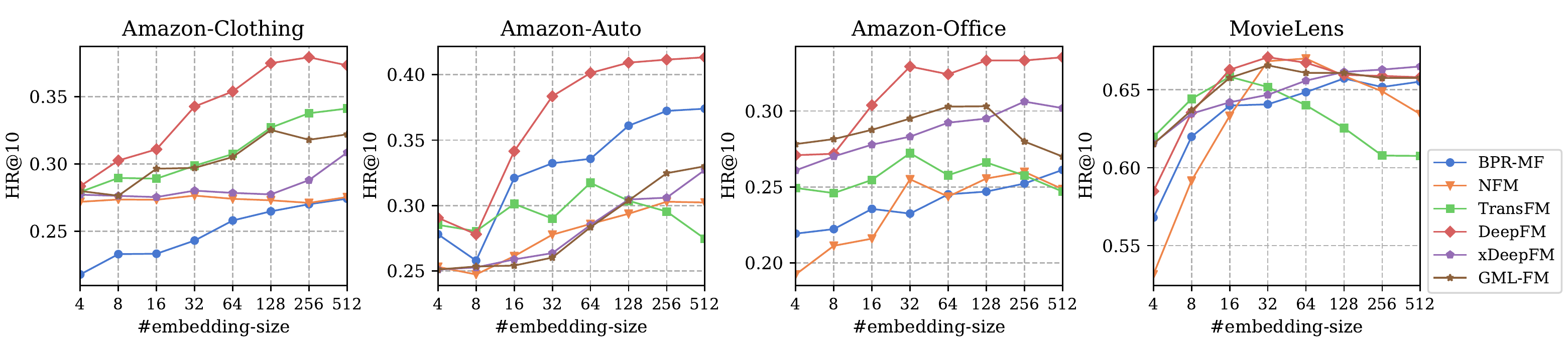}
  \caption{The influence of the embedding size on all the baselines and our method with respect to HR@10 on the top-n recommendation task.}\label{fig:embed_t}
\end{figure*}

\begin{itemize}
\item The introduced transformation weight is very critical for enhancing the model's representation capability. In particular, on the Mercari-Ticket dataset, the absolute improvement of HR is 35.46\% and 49.26\% for Euclidean and Mahalanobis distance functions on the top-n recommendation task, respectively. Notice that the values in the row of `\#layers 0' are the results of the conventional Euclidean distance with the transformation weight. In view of this, simply introducing intra-attribute interactions to the plain Euclidean distance variant can significantly boost the model performance, which evidently highlights the importance of intra-attribute interaction modeling and the effectiveness of our method.
\item The performance of the Mahalanobis distance function is inferior to that of the conventional Euclidean one when the transformation weight is removed. However, with the introduction of the transformation weight, the Mahalanobis method can consistently outperform the Euclidean one. The reason behind this is that, with the transformation weight, the Mahalanobis distance method is more suitable for capturing the inherent correlations between features than the simple Euclidean distance does.
\item When the number of layers increases from 0 to 2, the performance is consistently improved. Nevertheless, when the number of layers is set to 3, a large deterioration can be observed. This is mainly because that more parameters lead to the over-fitting problem. It can be seen from this table that two layers of deep neural networks is a reasonable choice on most occasions.
\item From the results of different distance functions, we found that the employed Euclidean one outperforms the other three variants consistently. Besides, the Manhattan and Chebyshev functions surpass the Cosine one, which is mainly because that the Cosine distance is being computed in a inner product fashion.
\end{itemize}

\subsection{Effects of Embedding Size (RQ3)}
To analyze the effect of different embedding sizes of our proposed method and the other baselines, we present the results of these methods with increasing embedding size on four datasets: Amazon-Clothing, Amazon-Auto,Amazon-Office and MovieLens. Note that we used the GML-FM with deep neural networks in this experiment. Figure~\ref{fig:embed_t} show the performance changes with the increasing embedding size on the top-n recommendation task. It can be observed that, for almost all the methods, with the increasing embedding size, the performance improves firstly, and then starts to converge or deteriorate. In general, models with smaller embedding sizes lack  the representation ability, while too large embedding sizes could lead to over-fitting.

From the figure, we can observe that our method can surpass all the baselines under most embedding sizes with a large margin, except for the NFM (i.e., the green one) on the MovieLens dataset. It demonstrates that the feature-level interactions inside attributes are important for improving recommendation performance under sparse settings (as the MovieLens dataset is the most dense dataset among all the experimented datasets and NFM does not take feature-level interactions into consideration). This is more practical since the cold-start and data sparsity problems are the main obstacle of nowadays recommender systems.

Another observation is that the performance of GML-FM is more stable with the increase of embedding size compared to other methods. Besides, GML-FM is less prone to over-fit when the embedding size grows larger on all datasets. This implicitly shows the robustness of GML-FM over other baselines.

\subsection{Attribute Effect Exploration (RQ4)}
As the newly collected Mercari dataset contains rich types of side information, we conducted this study to explore the effect of the side information on two sub-datasets on the top-n recommendation task. The results of our proposed method with different attributes are shown in Table~\ref{tab:mercari}. The detailed attributes are: `base' refers to user and item, `cty' refers to item category, `cdn' refers to item condition (e.g., 70\% new), and `shp' refers to shipping information (e.g., shipping duration: 2 days, shipping method: air flight). Note that the item condition and shipping information is unique in our collected dataset.

From Table~\ref{tab:mercari}, we can observe that the method without any side information performs unsatisfactorily (38.29\% and 29.52\% absolute degradation of HR on the Mercari-Ticket and Mercari-Books compared to the model with all attributes, respectively). After considering the item category attribute in our method, a large improvement can be observed. Moreover, with the additional information of the item condition, the performance degrades slightly. In contrast, with item shipping information, the performance is improved. This indicates that the information of product condition does not provide discriminative features for purchasing prediction. In contrast, the shipping method is strongly related to the shipping duration and costs, which are more important for users. Finally, it is interesting to find that with additional attributes of both the condition and shipping method (i.e., all the contextual information), the performance can be further improved. It also indicates that with more side information, the performance can be improved in general. The results well demonstrate the complex interactions between all features.

\begin{table}
  \caption{The influence of different attributes of two Mercari sub-datasets on our proposed method on the top-n recommendation setting.}\label{tab:mercari}
  \centering
  \scalebox{1.0}{
  \begin{tabular}{|l|cc|cc|}
    \hline

                        & \multicolumn{2}{c|}{Mercari-Ticket}   & \multicolumn{2}{c|}{Mercari-Books}\\
    \hline
    Attributes          & HR        & NDCG                      & HR            & NDCG              \\
    \hline
    \hline
    base                & 0.1953    & 0.1028                    & 0.1506        & 0.0674            \\
    base+cty            & 0.5501    & 0.2580                    & 0.4430        & 0.2094            \\
    base+cty+cdn        & 0.5323    & 0.2483                    & 0.4457        & 0.2102            \\
    base+cty+shp        & 0.5645    & 0.2777                    & 0.4465        & 0.2130            \\
    base+all            & 0.5782    & 0.2894                    & 0.4458        & 0.2143            \\
    \hline
  \end{tabular}}
\end{table}

\subsection{Cold-Start Scenario (RQ5)}
\begin{figure}
  \centering
  \includegraphics[width=\linewidth]{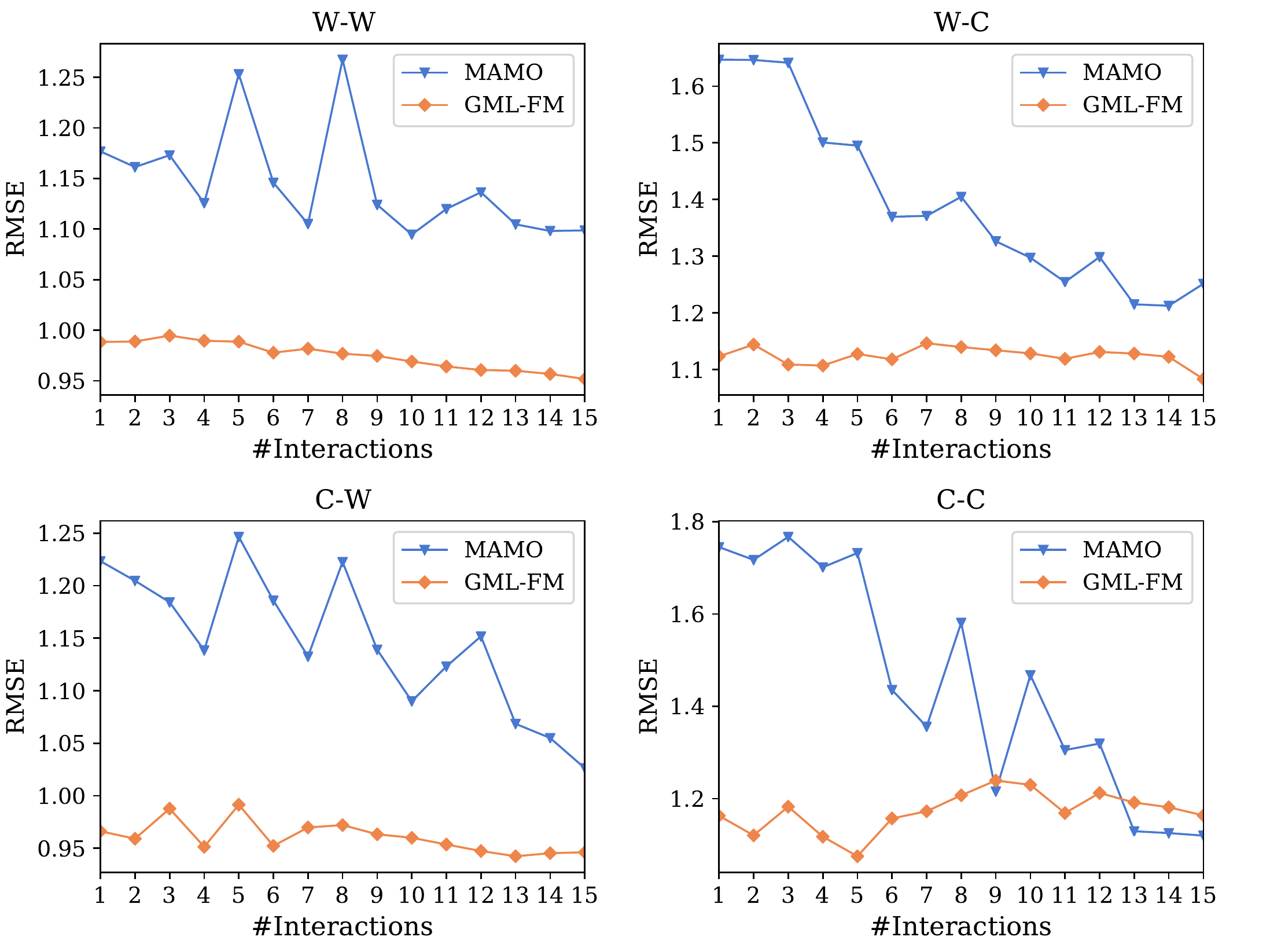}
  \caption{Performance comparison on cold-start rating prediction settings over the MovieLens dataset. The \#Interactions denotes the number of interactions in the training set for tested users.}\label{fig:cold}
\end{figure}

We conducted experiments to explore whether the proposed method is effective or not under cold-start settings. Towards this end, we compared it with a strong baseline - MAMO~\cite{mamo}, which adopts meta-learning to specially tackle the cold-start issue in recommendation. MAMO designs two memory matrices to provide personalized initialization instead of learning a global sharing initialization parameters for all users as previous meta-learning-based recommendation methods did. We adopted the official implementation\footnote{https://github.com/dongmanqing/Code-for-MAMO.} and analyzed the performance on four scenarios: 1) existing users for existing items (W-W) ; 2) existing users for cold items (W-C); 3) cold users for existing items (C-W); and 4) cold users for cold items (C-C). We leveraged the exactly same setting with ~\cite{mamo} to group users into warm and cold ones according to their first comment time; and group items into warm and cold ones based on the interacted times. The experiments are conducted on the MovieLens and the results are illustrated in Figure~\ref{fig:cold}. We can observe that:
\begin{itemize}
    \item Surprisingly, our GML-FM can outperform MAMO consistently and significantly. And the performance gap is even larger under very sparse settings, i.e., when the number of interactions is less than 5. This inspires us to develop more practical FM models for overcoming the cold-start  problem in recommendation compared with the meta-learning techniques.
    \item In general, the recommendation performance is gradually enhanced with more interactions in the training set. It is reasonable since more interactions provide more evidence for model learning.
\end{itemize}

\subsection{Case Study (RQ6)}
\begin{figure}
  \centering
  \includegraphics[width=0.95\linewidth]{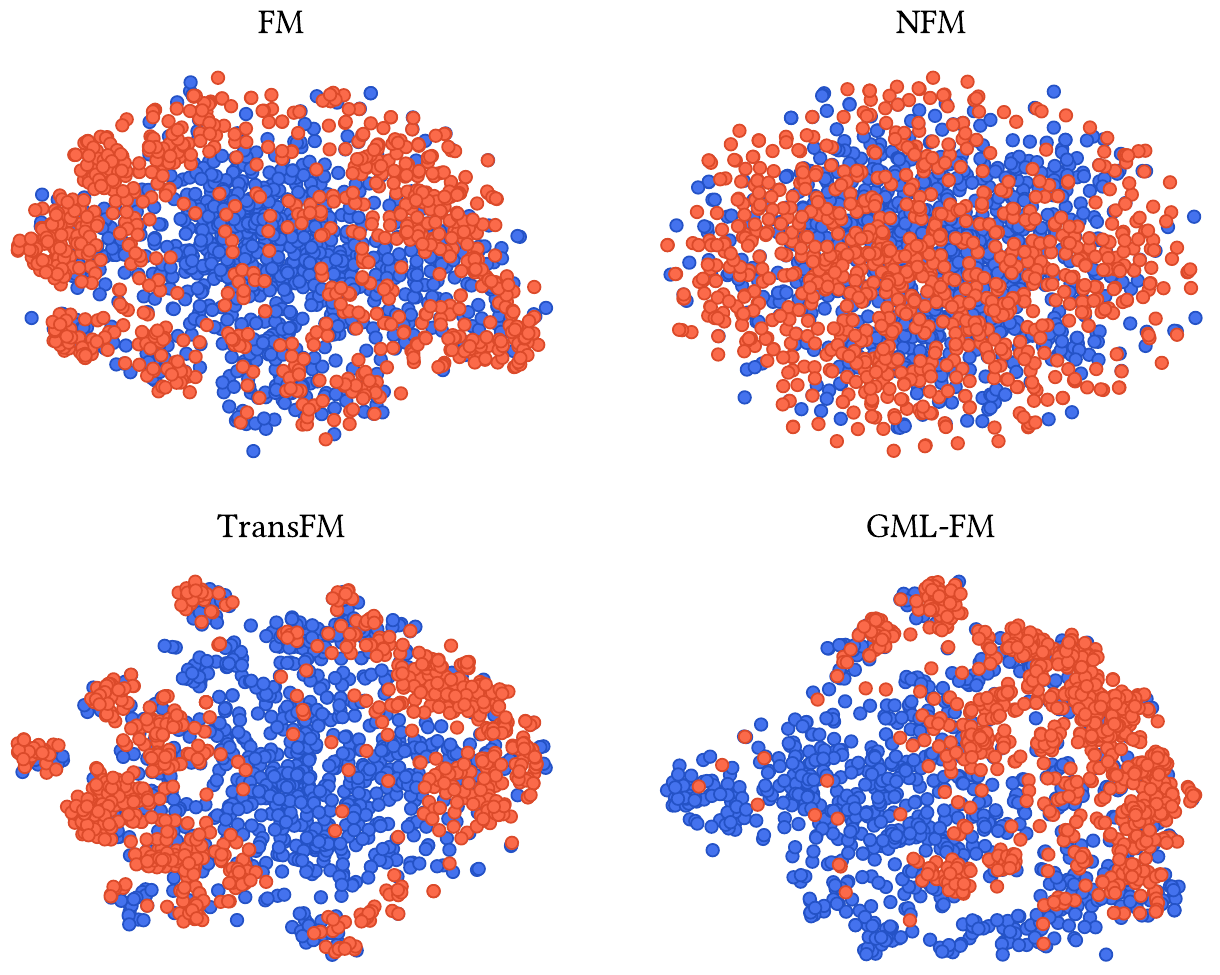}
  \caption{T-SNE visualization of item IDs embeddings of user ID 709 on four methods. The samples with red color represent the user interacted items (i.e., positive items), while the ones with blue color represent the user non-interacted ones (i.e., negative items).}\label{fig:tsne_1}
\end{figure}

\begin{figure}
  \centering
  \includegraphics[width=0.95\linewidth]{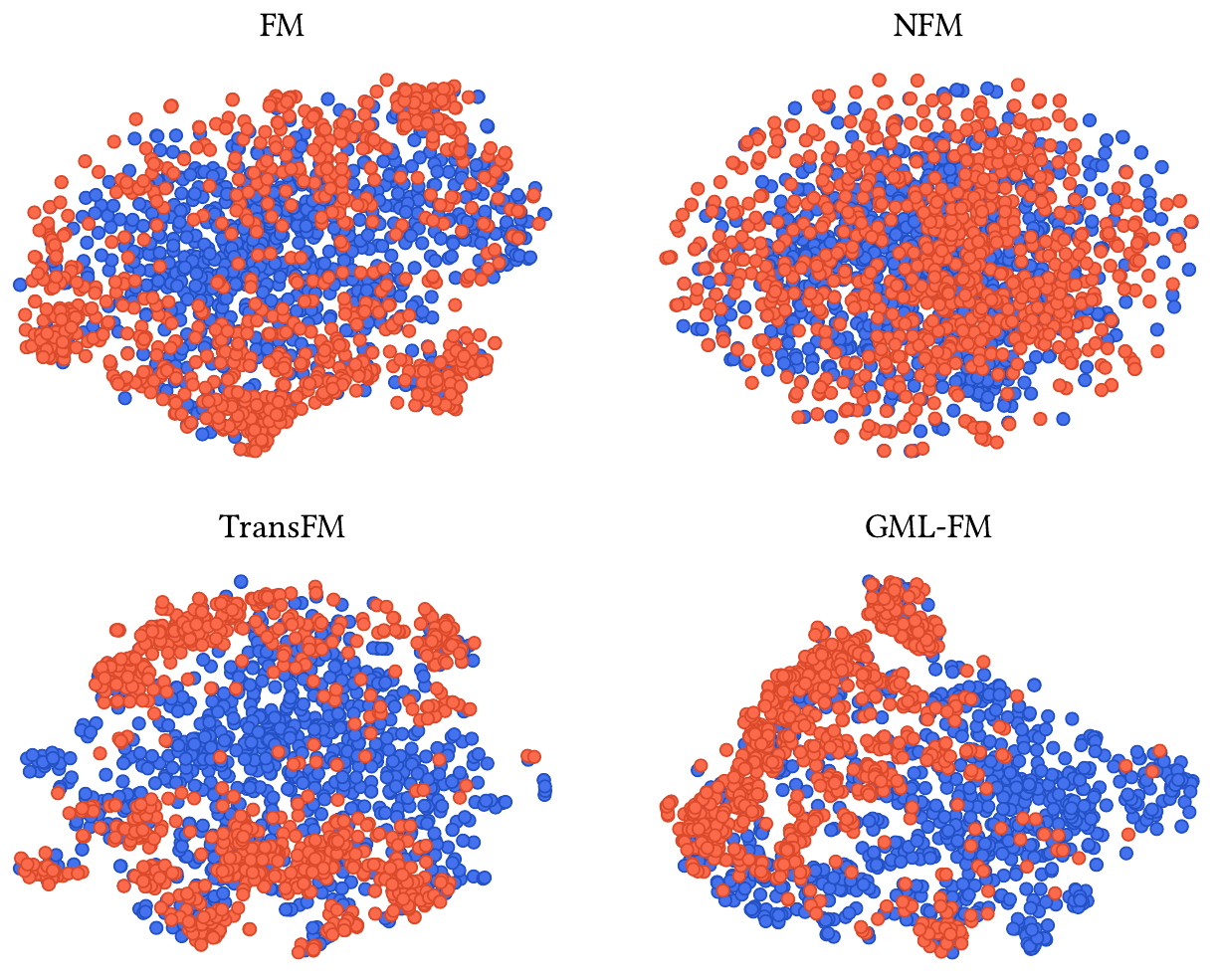}
  \caption{T-SNE visualization of item IDs embeddings of user ID 1050 on four methods. The samples with red color represent the user interacted items (i.e., positive items), while the ones with blue color represent the user non-interacted ones (i.e., negative items).}\label{fig:tsne_2}
\end{figure}

To intuitively understand what the proposed method and other FM-based baselines learn in the latent space, we visualize the item IDs embeddings of two users and illustrate them in Figure~\ref{fig:tsne_1} and Figure~\ref{fig:tsne_2}, respectively. In particular, we used two groups of item IDs embeddings and reduced the dimension to 2 for visualization. The two groups of items are: 1) positive samples: the items that user has interacted with in the training set, which are expressed with brown color; and 2) negative samples: the randomly sampled items that user has not interacted in the training set (equal quantity with the first group), which are expressed with blue color. As shown in Figure~\ref{fig:tsne_1} and Figure~\ref{fig:tsne_2}, the metric learning based methods (i.e., TransFM and GML-FM) demonstrate strong superiority over inner product ones (i.e., FM and NFM), where the positive samples under metric learning based methods are grouped into clusters, while the ones under inner product do not show obvious patterns. Compared to TransFM, the proposed GML-FM can significantly cluster the positive samples to one side. The reason why there is no specific borders between positive and negative samples is that we used the positive samples in the training set, while some negative ones may have interactions with the user in the testing set. Concretely, for user ID 709 in Figure~\ref{fig:tsne_1}, the GML-FM clustered positive items are mainly on the right side, and for user ID 1050 in Figure~\ref{fig:tsne_2}, the GML-FM clustered positive items are on the top-left areas. Compared with TransFM, our GML-FM can learn more expressive and effective attribute representations since the positive item embeddings themselves for one user should be more similar than those with negative ones. This also demonstrates that the feature interactions inside attributes are definitely important for learning better attribute representations and the GML-FM can effectively capture these interactions and surpass that of TransFM.

\section{Related Work}\label{related_work}
\subsection{Factorization Machines}
Factorization Machines (FMs) are a general model class working with any real valued feature vectors. Different from SVMs, FMs model all attribute interactions using factorized parameters, which enables them to estimate interactions even in tasks with huge sparsity where SVMs fail. FMs can be generalized to a variety of famous models, including SVMs~\cite{svm}, MFs~\cite{mf}, PITF~\cite{pitf} and FPMC~\cite{fpmc}.

Different from the MF-based methods (e.g., BPR~\cite{bpr} and the recent graph-based ones NGCF~\cite{ngcf}, LightGCN~\cite{lightgcn} or GIN~\cite{gin}), FMs can naturally consume rich contextual information (namely attributes) into recommendation, such as item category~\cite{deepfm,xdeepfm} or user mood when watching movies~\cite{context}. One notable merit regrading attribute exploitation of FMs over other methods~\cite{wide,din} can be attributed to the modeling of low- and high-order attribute interactions~\cite{deepfm,xdeepfm}.  However, traditional FMs usually take all attribute interactions equally, which is not reasonable since interactions are not equally important and some interactions may even introduce noises to the score prediction. Several studies have devoted to solving this problem by leveraging gradient boosting~\cite{gradientfm}, attention mechanisms~\cite{afm} and Bayesian personalized feature interaction~\cite{bayesian}. For example, Xiao et al.~\cite{afm} presented an attention-based FM to automatically learn the importance of different attribute interactions for the final score prediction. The learned attention weights guide FMs to treat different interactions discriminately. Besides, in order to model the non-linear and complex second-order attribute interactions, He et al.~\cite{nfm} introduced deep neural networks into FMs, where a Bi-Interaction is firstly operated on attribute interactions, followed by several fully connected deep layers.

In addition, FMs have been developed and extended to many different tasks~\cite{poly,multilinear,ctr}. For instance, Lu et al.~\cite{multilinear} proposed an efficient multi-linear FM model to address the multi-task multi-view problem. Petroni t al.~\cite{relationfm} developed a novel MF model based upon FMs for open relation extraction, which can effectively integrate various side information such as metadata information. Effective yet efficient higher-order interactions are also studied in~\cite{poly,higher}.

To the best of our knowledge, previous FM-based methods focus on the attribute-level interactions (i.e., inter-attribute interactions), while the feature-level interactions inside attributes (i.e., intra-attribute interactions) are leaving untapped. However, only modeling the attribute interactions is insufficient to capture finer grained feature interactions, resulting in sub-optimal performance. In this work, we attempt to bridge this gap by proposing a generalized metric learning based FMs method to model the feature interactions of an attribute.

\subsection{Metric Learning}
Metric learning has attracted more and more attention over the past few years. Various applications utilize it to learn an appropriate distance metric for specific decision making. For example, ~\cite{clusterml} employed semi-definite programming to learn a Mahalanobis distance metric for clustering. Different from clustering where all the objects with same labels are grouped together, \cite{metriclearning,knnml} designed an efficient distance metric learning algorithm to make the object and its k-nearest neighbors close in the mapped feature space. One typical way to implement this is to automatically learn a distance metric that pulls similar pairs together and pushes dissimilar pairs apart.

Recently, several efforts have been dedicated to combining metric learning with recommendation such as collaborative filtering~\cite{cf}. For instance, Bachrach et al.~\cite{speed} present an order preserving transformation, mapping the maximal inner product search problem into an Euclidean nearest neighbor search one. Hsieh et al.~\cite{cml} proposed a collaborative metric learning method to connect the metric learning and collaborative filtering. This method encodes user-item relations and user-user/item-item similarity into a joint metric space, and provides a suite of feature fusion techniques to make it feasible. This idea is further extended by LRML~\cite{relationml}, modeling the latent relations that describe each user-item interaction, and MAML~\cite{diverse}, considering user's varying preferences on items.

As metric learning shows dominant advantage over inner product in recommendation, Pasricha et al.~\cite{transfm} recently adapted it to model side information with FMs. Specifically, the Euclidean distance function is employed to estimate the distance between the two attribute embedding vectors, where one of them is the addition of the embedding vector and a translation vector of the current attribute. Different from the work introduced in~\cite{transfm}, we exploited metric learning techniques in a novel way, which then is leveraged to model the feature interactions inside attributes. 
\section{Conclusion and future work}\label{conclusion}
In this work, we present a novel FM framework equipped with generalized metric learning techniques, namely GML-FM, to model the finer grained feature correlations in FMs. More concretely, we present two methods under the framework, GML-FM$_{md}$ and GML-FM$_{dnn}$. The former one adopts the Mahalanobis distance function which contains a learnable semi-positive matrix. It is able to capture the linear correlations between features. The latter one utilizes the deep neural networks to capture the non-linear feature correlations. Furthermore, we  designea transformation weight, which can extend the values of metric learning based FMs to cover the whole real number space and thereby increase the representation capability. In addition, we further propose an efficient approach to reducing the computation complexity. Another contribution is that we collect a new large-scale second-hand trading dataset to facilitate the study of cold-start and data sparsity problems in recommender systems. Extensive experiments on several benchmark datasets and the newly developed dataset validate the effectiveness of the proposed method.

In the future, we will explore pair-wise learning technique for GML-FM by enhancing GML-FM with the Bayesian Personalized Ranking (BPR) approach. Furthermore, as the Mercari dataset provides rich user-submitted queries and the corresponding clicking information, we will adapt and apply the GML-FM method to product search tasks to explore the effectiveness of the proposed method in other domains. 
\appendices
\section{Detailed Derivation for Equation 10} \label{appendix}

\begin{equation*}\label{equ:simfm_mdis1}
\begin{aligned}
  f(\bm{x}) &= \sum_{i=1}^{n} \sum_{j=i+1}^{n} \bm{h}^{T} (\bm{v}_i \odot \bm{v}_j) D(\bm{v}_i, \bm{v}_j) x_i x_j, \\
            &= \sum_{i=1}^{n} \sum_{j=i+1}^{n} \bm{h}^{T} (\bm{v}_i \odot \bm{v}_j) (\bm{v}_i - \bm{v}_j)^T \bm{M} (\bm{v}_i - \bm{v}_j)x_i x_j, \\
            &= \frac{1}{2} \sum_{i=1}^{n} \sum_{j=1}^{n} (\bm{v}_i^T diag(\bm{h}) \bm{v}_j \bm{v}_i^T \bm{M} \bm{v}_i -  \\
            & \: \: \: \: \: \:  2\bm{v}_i^T diag(\bm{h}) \bm{v}_j \bm{v}_i^T \bm{M} \bm{v}_j + \bm{v}_i^T diag(\bm{h}) \bm{v}_j \bm{v}_j^T \bm{M} \bm{v}_j) x_i x_j, \\
            &= \frac{1}{2} \sum_{i=1}^{n} \sum_{j=1}^{n} (\bm{v}_j^T diag(\bm{h}) \bm{v}_i \bm{v}_i^T \bm{M} \bm{v}_i -  \\
            & \: \: \: \: \: \: 2\bm{v}_j^T diag(\bm{h}) \bm{v}_i \bm{v}_i^T \bm{M} \bm{v}_j + \bm{v}_i^T diag(\bm{h}) \bm{v}_j \bm{v}_j^T \bm{M} \bm{v}_j) x_i x_j, \\
            &= \frac{1}{2} \sum_{j=1}^{n} \bm{v}_j^T x_j \sum_{i=1}^{n} diag(\bm{h}) \bm{v}_i \bm{v}_i^T \bm{M} \bm{v}_i x_i - \\
            & \: \: \: \: \: \:  \sum_{j=1}^{n} \bm{v}_j^T diag(\bm{h}) ( \sum_{i=1}^{n} \bm{v}_i \bm{v}_i^T x_i) \bm{M} \bm{v}_j x_j + \\
            & \: \: \: \: \: \:  \frac{1}{2} \sum_{i=1}^{n} \bm{v}_i^T x_i \sum_{j=1}^{n} diag(\bm{h}) \bm{v}_j \bm{v}_j^T \bm{M} \bm{v}_j x_j, \\
            &= \sum_{j=1}^{n} \bm{v}_j^T x_j \sum_{i=1}^{n} diag(\bm{h}) \bm{v}_i \bm{v}_i^T \bm{M} \bm{v}_i x_i - \\
            & \: \: \: \: \:   \sum_{j=1}^{n} \bm{v}_j^T diag(\bm{h}) ( \sum_{i=1}^{n} \bm{v}_i \bm{v}_i^T x_i) \bm{M} \bm{v}_j x_j, \\
\end{aligned}
\end{equation*}

%
\ifCLASSOPTIONcompsoc
  \section*{Acknowledgments}
This work is supported by the National Natural Science Foundation of China, No.:61902223, No.:61772310, No.:U1936203; the Shandong Provincial Natural Science and Foundation, No.: ZR2019JQ23, the Innovation Teams in Colleges and Universities in Jinan, No.:2018GXRC014.

\else
  \section*{Acknowledgment}
\fi


\ifCLASSOPTIONcaptionsoff
  \newpage
\fi

\bibliographystyle{IEEEtran}
\bibliography{tkde19}

\end{document}